\documentclass{osa-article}

\journal{osajournal}

\articletype{Research Article}

\usepackage{lineno}
\usepackage{ulem}

\begin{document}

\title{Real-time two-photon interference from distinct molecules on the same chip}

\author{Rocco Duquennoy\authormark{1,2,3}, Maja Colautti\authormark{2,3,4}, Ramin Emadi\authormark{1,2}, Prosenjit Majumder \authormark{2,4}, Pietro Lombardi\authormark{2,4,5}, and Costanza Toninelli\authormark{2,4}}

\address{

\authormark{1}Physics Department - University of Naples, via Cinthia 21, Fuorigrotta 80126, Italy\\
\authormark{2}National Institute of Optics (CNR-INO), Via Nello Carrara 1, Sesto F.no 50019, Italy\\
\authormark{3}Co-first authors with equal contribution\\
\authormark{4}European Laboratory for Non-Linear Spectroscopy (LENS), Via Nello Carrara 1, Sesto F.no 50019, Italy \\
\authormark{5}lombardi@lens.unifi.it \\

}




\begin{abstract}



Scalability and miniaturization are hallmarks of solid-state platforms for photonic quantum technologies. Still a main challenge is two-photon interference from distinct emitters on chip. This requires local tuning, integration and novel approaches to understand and tame noise processes. A promising platform is that of molecular single photon sources. Thousands of molecules with optically tuneable emission frequency can be easily isolated in solid matrices and triggered with pulsed excitation. We here discuss Hong-Ou-Mandel interference experiments using several couples of molecules within few tens of microns. Quantum interference is observed in real time, enabling the analysis of local environment effects at different time-scales. 
\end{abstract}


\section{Introduction}
Quantum optics and photonics have brought amongst the most advanced applications in quantum technologies, especially in the fields of communication \cite{Kolodynski2020, Yin2020}, sensing and metrology \cite{giovannetti11, Ansari2021}, computing \cite{OBrien2009,Arrazola2021} and simulation \cite{Aspuru-Guzik2012, Sparrow2018, HanSen2020}. Also distributed sensing \cite{Guo2020} and quantum networks \cite{Wehner2018, Lodahl2015, Azuma2015} all require efficient links to quantum states of light. Interestingly, in many of these contexts, two-photon interference (TPI) - an utterly quantum effect- stands as a fundamental process\cite{Bouchard2021}.

In terms of physical systems, solid-state platforms of different kinds are routinely exploited and provide key components in quantum photonics, from non-classical light sources \cite{Aharonovich2016, Lounis2005, Arrazola2021}, to quantum-circuits \cite{Politi2008, Crespi2013}, -sensors \cite{Jackson2021}, -memories \cite{LagoRivera2021} and detectors \cite{Pernice2012}, everything potentially miniaturized in integrated optical chips \cite{Politi2008, Lodahl2015, Babin2021, Sipahigil2016, Lenzini2018, Toninelli2021}. 
However, the interaction of the electromagnetic field with the host material, and of the different physical elements within it, strongly affects the quality of such quantum resources. In particular, the first order coherence of single-photon wavepackets generated by quantum emitters in the solid state strongly reflects the environmental noise around them, degrading the visibility of TPI \cite{Jantzen2016, Kambs2018}. Looking at it from a different point of view, TPI might become an exquisite tool to investigate dynamical processes at the emitter place, by measuring joint spectral properties of photon pairs \cite{Bouchard2021,Cimini2021a}. 
Considering single-photon pulses from distinct emitters in particular, besides dephasing, quantum interference is sensitive to frequency fluctuations occurring on the minute-long time scale of the measurement. Indeed, even highly indistinguishable single photon sources such as those based on epitaxial quantum dots\cite{Tomm2021, Somaschi2016, Uppu2021}, color centers in diamond\cite{Aharonovich2016}, or isolated molecules \cite{trebbia2010indistinguishable, Rezai2018, Lombardi2021}, have shown a lower interference visibility when measured against each other \cite{Reindl2017, Lettow2010a, Sipahigil2014, Sipahigil2012, Bernien2012, Weber2019}. Widely used strategies to counteract visibility losses are spectral filtering, temporal post-selection and coherent excitation schemes, all of which entail lower-brightness sources \cite{Stockill2017, He2013, Gao2012}. A record-high value was recently demonstrated for gated GaAs quantum dots hosted in separated wafers and cryostats \cite{Zhai2021}. However, to unlock the full potential of solid state quantum emitters for a quantum advantage in photonics \cite{HanSen2020}, tunable sources of indistinguishable photons should be ideally available in a single chip. The opportunity of scaling up the number of single photon states by adding more quantum emitters on the same platform, is still mostly wishful thinking and current approaches are limited to temporal demultiplexing from a single emitter \cite{Wang2019b, Loredo2017}, or multiple probabilistic sources \cite{Paesani2019}. 

Molecular quantum emitters have already shown a great potential for integration in photonic circuits \cite{Toninelli2021, Colautti2020, Lombardi2017a, Turschmann2017, Grandi2019, Boissier2021, Hail2019}. In this paper we study the problem of TPI from distinct molecular emitters on chip, attaining and combining together the following milestones: simultaneously addressing on the same sample several single-molecules operating as on-demand single photon sources,  tuning independently their relative zero-phonon line (ZPL) frequency, measuring in semi real-time two photon interference from such distinct sources and extracting information about joint properties of the photon pairs. 

\begin{figure}[h!]
\centering\includegraphics[width=10cm]{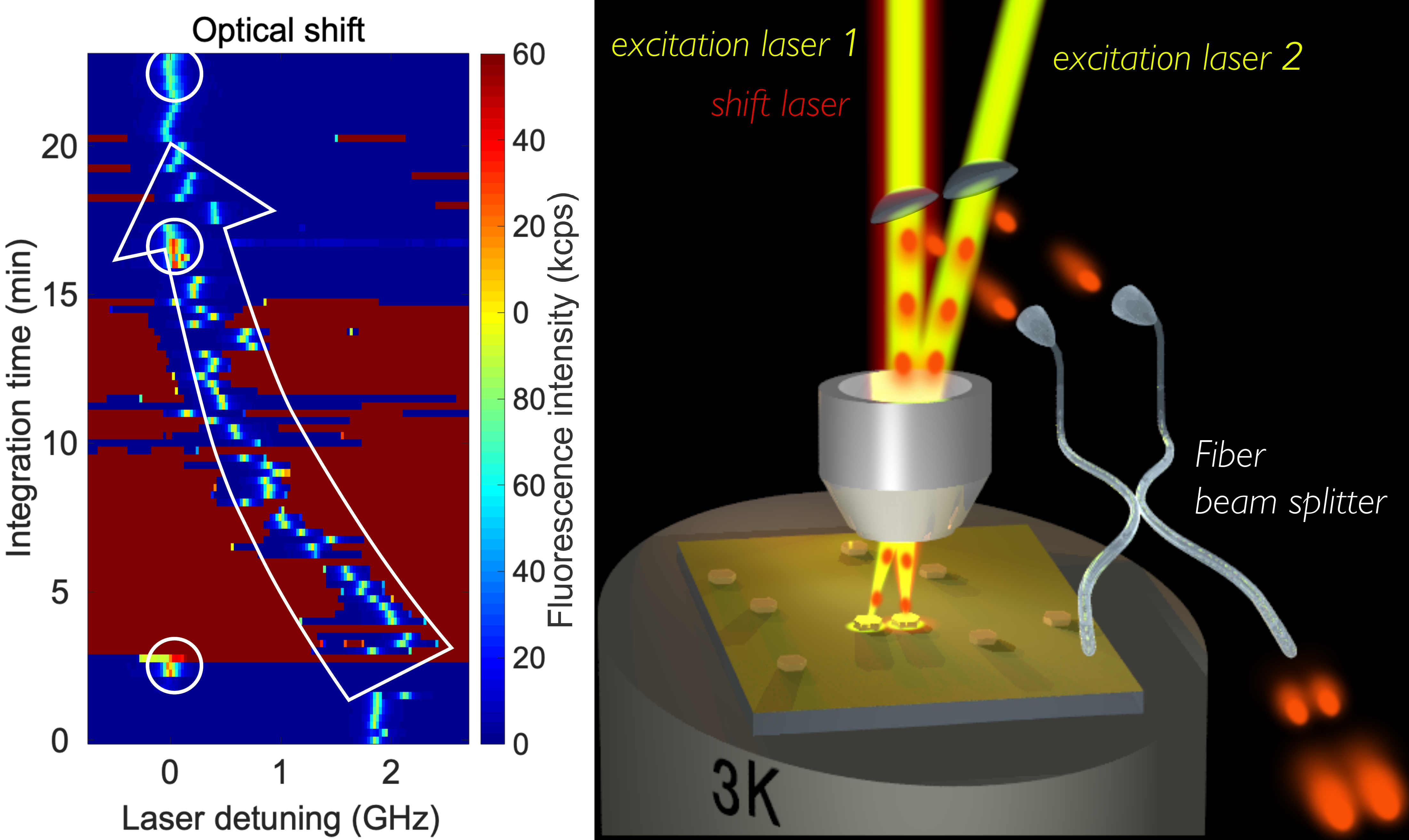}

\caption{\textbf{Artistic view of the experiment and working principle:} The zero phonon lines of molecules embedded in different nanocrystals on the same gold-coated chip are alternatively shifted by means of a focused laser, called shift laser. By looking at their excitation spectrum after several shifting periods, corresponding to red-saturated lines in the figure left panel, the two spectra are overlapped with a peak frequency difference of about $50\,$MHz. The three circles highlight the first molecule excitation spectrum, while the arrow is a visual guide to follow the second molecule excitation spectrum during the shifting sequence, until it matches the first molecule ZPL. Once the molecules' ZP emission lines are tuned to resonance, their emission is spatially separated and then recombined on a fiber beam splitter. When also polarization, spatial mode and arrival time are matched, two-photon interference is observed in the coalescence of photons at the output ports.}\label{fig1}
\end{figure}
\section{Concept and Experimental Setup}
For this work we consider a sample of dibenzoterrylene (DBT) molecules embedded in anthracene nanocrystals\cite{Pazzagli2018}. These are dispersed on a gold-coated silica substrate in order to improve collection efficiency and covered with a 50 nm-thick PVA layer, for protection against matrix sublimation. On a similar system, high collection efficiency\cite{lombardi2020molecule} and Hong-Ou Mandel interference under pulsed excitation\cite{Lombardi2021} were recently demonstrated, in the case of photons generated by a single DBT molecule. This time, couples of molecules are selected, whose emission is close  both in spectrum (ZPLs less than 100 GHz apart) and in space (typically within a distance of $30\,\mu m$). Thanks to a dual confocal excitation setup, the two molecules can be addressed at the same time, falling within the same objective field of view. After a long pass filter that rejects the scattered pump light and a notch filter isolating the ZPL component, the collected fluorescence from each molecule is spatially separated by means of a D-shaped pick-off mirror positioned in an intermediate image plane, and conveyed in two distinct fiber-coupled paths. Further manipulation to probe the properties of the collected photons is done with fiber-based elements and single photon detectors (for an artistic description see Fig.\ref{fig1}). Molecules are interrogated with a CW laser which can be scanned around the ZPL molecular resonance around 784 nm, and with a pulsed laser at 766 nm for the triggered non-resonant pumping.
The experiment is made possible by the recently discovered effect of laser-induced charge-separated state formation in molecular matrices, giving rise to a local Stark shift for the DBT molecules therein embedded \cite{Colautti2020b}. To this aim, an additional pump beam, called shift beam, is aligned to the optical path, with an intensity typically one order of magnitude higher than that of the trigger excitation laser. The frequency shift appears linear with the applied dose for short times and saturates after hour-long illumination, for an overall achievable shift of the order of $50-100\;GHz$. Characteristic frequency traces of two selected molecules acquired during dark times between illumination periods is shown in Fig.\ref{fig1}. Color coded is the emission intensity in the phonon wing, while the resonant laser is scanned through the ZPL of one or the other molecule. The spectra collected in each row have a total integration time of $15\,$s, and are accumulated over a total illumination time of about $25\,$min, with a power varying from $10$ to $60\;\mu W$ for the shifting laser and amounting to $1\; nW$ in the resonant scanning beam. By moving one of the two excitation spots, we can selectively shift and/or probe one of the two molecule of the couple. In this way, iterating shift and probe intervals in sequence (red saturated and blue background in the figure, respectively), we bring two molecules on resonance within a frequency difference of about $50\,$MHz (see Fig.\ref{fig1}), smaller than  the average molecules' linewidth.

\section{Results and Discussion}

\begin{figure}[h!]
\centering\includegraphics[width=13cm]{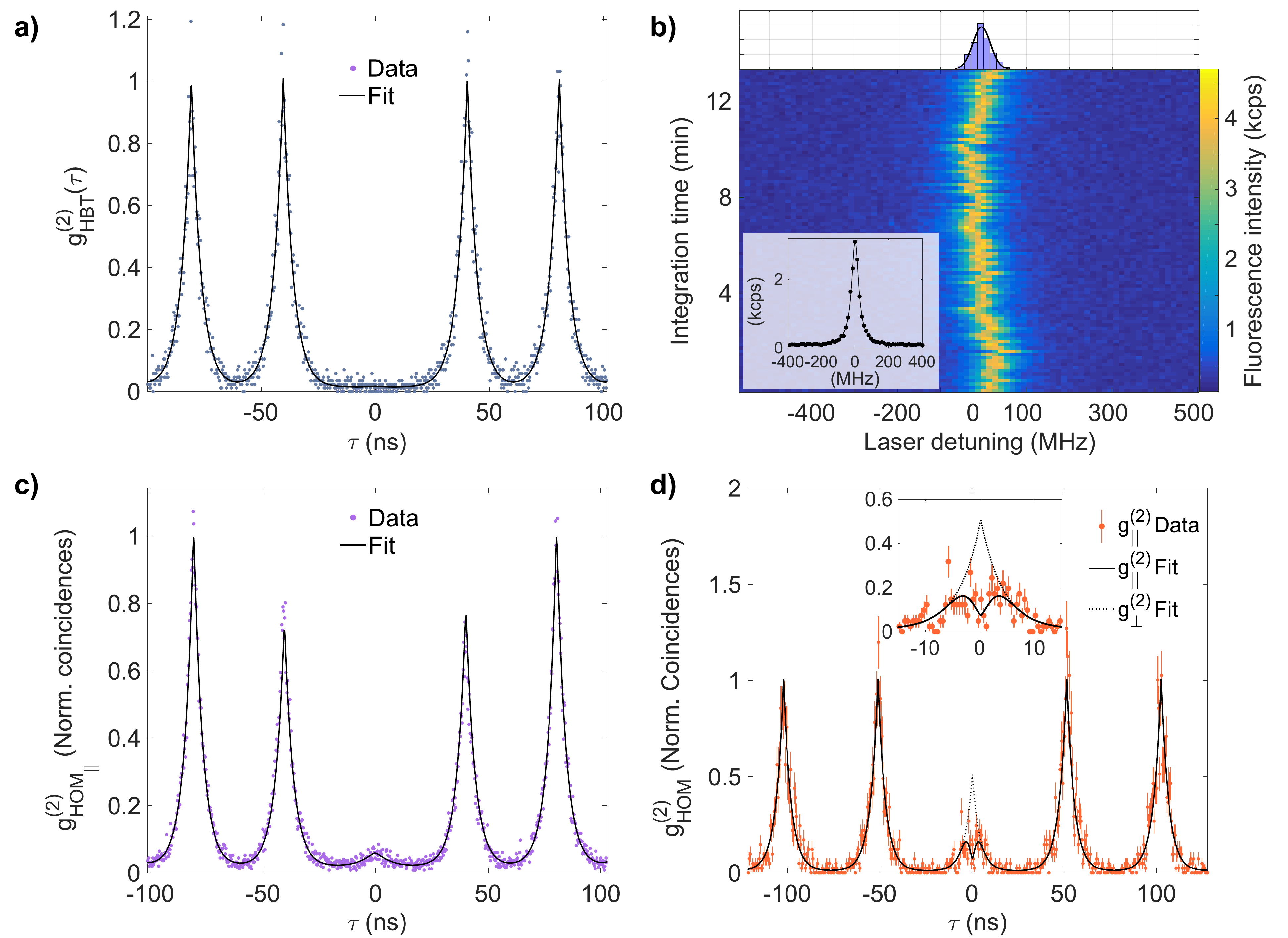}

\caption{\textbf{Emission properties and two-photon interference:} \textbf{a)} The single-photon-state purity is estimated from the second order correlation function obtained detecting coincidences in a Hambury-Brown and Twiss configuration, under pulsed operation. $g_{HBT}^{(2)}[0]=0.003 \pm 0.013$ is obtained, normalizing the area at zero delay with the average peak area for longer delays. \textbf{b)} Fluorescence intensity as a function of the excitation laser frequency. In this case, statistics on the width and on the positions of the signal peak provides an estimation of its linewidth and spectral wondering, yielding $52\pm3\;$MHz and $\sim 20\;$MHz, respectively. \textbf{c)} Indistinguishability between photons emitted by the same molecule is probed via Hong-Ou-Mandel interference (parallel polarizations case). Best fit to the data yields $g_{HOM}^{(2)}[0]=0.06\pm0.015$. \textbf{d)} Indistinguishability between photons emitted by two separated molecules, probed via Hong-Ou-Mandel detection setup (parallel polarizations case). Acquisition is performed in time-tagged mode and the data are integrated over 30 seconds, yielding $V=(40\pm7)\%$ and $v= 96\%\pm 8\%$. The curve for orthogonal polarizations is calculated considering no interference term in Eq.\ref{HOM}.
The coincidences histograms are obtained with a repetition rate of $25\;MHz$ and $20\;MHz$ for panel (d), integrating the following  photon flux per SPAD: c) 20kcps for 6min; b) 15kcps for 6min; d) 20kcps for 3min.}\label{fig2}
\end{figure}

Before studying two-photon interference, molecules have been characterized by measuring the second order autocorrelation function $g^{(2)}_{HBT}(\tau)$ in the Hanbury-Brown and Twiss (HBT) configuration, the ZPL excitation spectrum, its spectral diffusion and single-molecule HOM autocorrelation $g^{(2)}_{HOM}(\tau)$. Typical results are shown in Fig.\ref{fig2}(a), (b) and (c), respectively. In particular, we observe that the normalized area of the residual peak at zero time delay of the HBT measurements in panel Fig.\ref{fig2}(a) amounts to only $g^{(2)}_{HBT}[0]=0.003 \pm 0.013$, where $g^{(2)}[x]$ represents the area of the peak around $x$ time delay, integrated over the pulsed-laser period, i.e. $40\,$ns. The purity of single photon emission is among the highest reported values.
Excitation spectroscopy allows instead to estimate the role of residual dephasing on the coherence of the electronic levels, and of their frequency fluctuations on long time scales. 

Fig.\ref{fig2}(b) shows in color scale the fluorescence intensity emitted in the phonon wing exhibiting a maximum when the resonant laser wavelength is tuned to the center of the ZPL. In the inset, a zoom-in of a typical scan is plotted. The  measurement is repeated under the very same conditions several times for a total amount of about $15\,$min. From such measurements, fitting different rows with lorentzian profiles (see inset) we obtain a statistical estimation for the line-width ($FWHM=52\pm3\,$MHz). Considering the total measurement time we can evaluate the spectral wondering and extrapolate a typical standard deviation for the distribution of the central emission frequency $\sigma \sim 20\,$MHz (see histogram on the panel top side). Considering the line-width value (or equivalently $T_2=1/(\pi FWHM)=(6.1\pm0.4)$\,ns)  and the excited state lifetime $T_1=4.0\pm 0.2$ $ns$, as derived from the best fit to the experimental HBT data, a characteristic dephasing time $T_2^*=(26\pm4)\,$ns can be estimated. It is important to remind here that the measurements displayed in this figure do not correspond all to the same molecule, as priority has been given to showing the clearest examples. The consistency in the analysis was fully discussed in our previous paper \cite{Lombardi2021}.

The results of the autocorrelation function for single-molecule data are reported in Fig.\ref{fig2}(c). The $g^{(2)}_{HOM}(\tau)$ is calculated by looking at coincidences between two single-photon detectors at the output ports of a symmetric beam splitter, after pulses from the same molecule are splitted in two arms and then recombined at the two input ports. The experimental results depicted in the panel are obtained when the polarization on both arms is set to parallel ($g^{(2)}_{HOM}[0]_{\parallel}$) and the laser period $T$ is adjusted to match the paths' relative delay, yielding maximum interference effects, with the following value for the visibility: $V=\frac{g^{(2)}_{HOM}[0]_{\bot}-g^{(2)}_{HOM}[0]_{\parallel}}{g^{(2)}_{HOM}[0]_{\bot}}= 0.88\pm0.03$, where $g^{(2)}_{HOM}[0]_{\bot}$ corresponds to the normalized counts in the zero-delay peak when orthogonal polarizations are set. The central peak of the curve is described by a function already introduced in Ref.\cite{Lombardi2021}, yielding as best-fit parameters: $T_1=4.0\pm 0.2$ $ns$ and  $FWHM=45\pm 3$\;MHz. 

The maximum expected visibility should hence be $T_2/2T_1=0.88\pm 0.14$ and is consistent with our estimation. Indeed, the $g_{HOM}^{(2)}(0)$, which accounts for all the experimental elements limiting interference, amounts to only $0.035\pm 0.02$ and corresponds to a post-selected visibility $(93\pm2)\%$. We note in passing that, comparing the amplitude of the neighbouring peaks at $\tau=\pm 2,3,4 T$ to the ones for longer time delays, there is no appreciable difference, which is a signature of negligible blinking effects. As a consequence, the visibility is equivalently estimated from now on-wards by normalizing to the average area of peaks other than the central one as follows:\begin{equation}
    V=\frac{g^{(2)}_{HOM}[nT]-2g^{(2)}_{HOM}[0]}{g^{(2)}_{HOM}[nT]} \;\;\;\;\;\;\;\;\;\;\;n>1
\end{equation}
\begin{figure}[h!]
\centering\includegraphics[width=13cm]{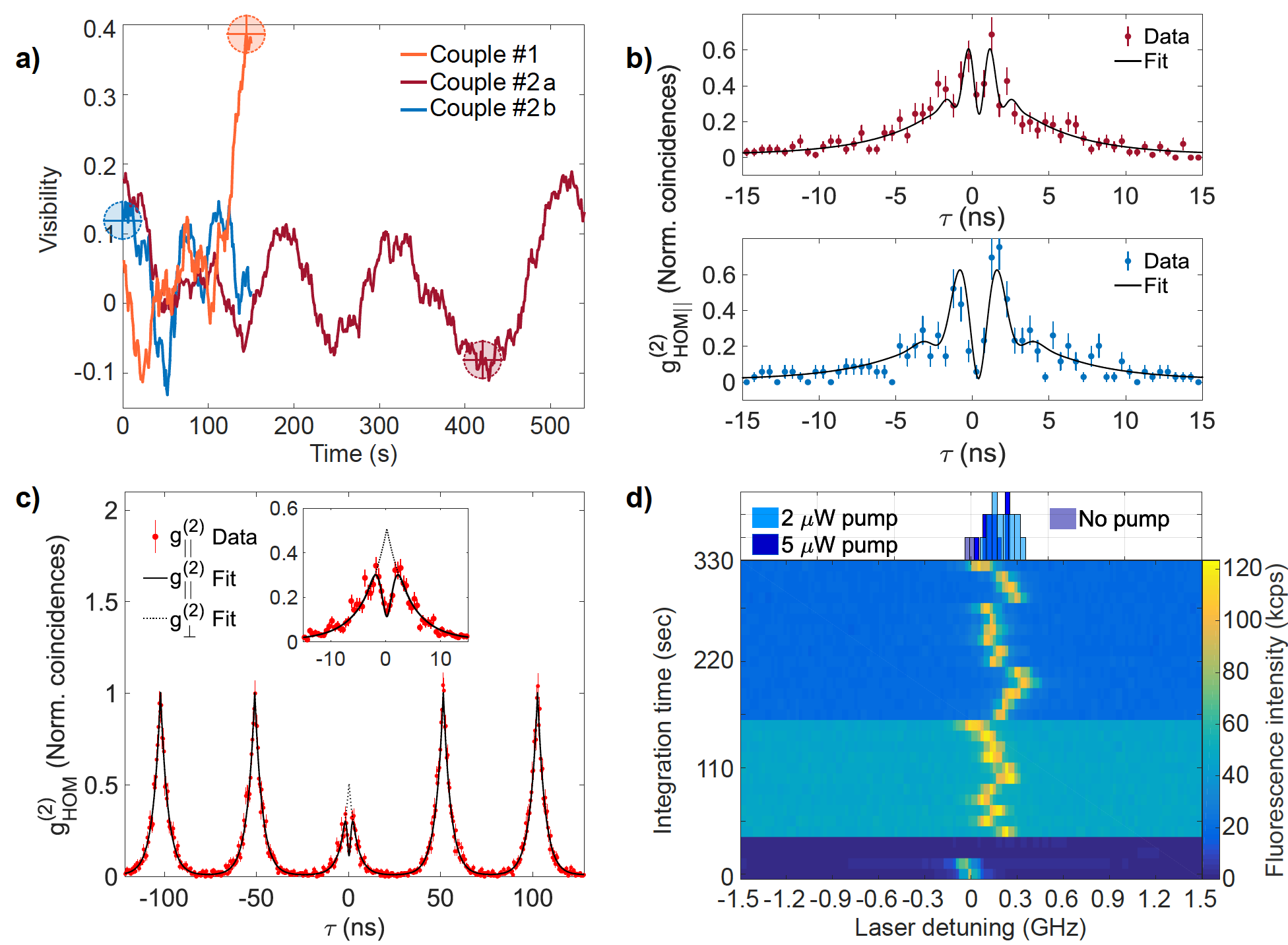}
\caption{\textbf{Two-photon interference from distinct molecules on chip:} \textbf{a)}  Temporal evolution of the TPI visibility calculated as in Eq. \ref{HOM}, shifting a $30$-s time window on data acquired in time-tagged mode, for two couples of molecules. \textbf{b)} $g_{HOM}^{(2)}(\tau)$ corresponding to selected visibility values plotted in panel a): color-coded and crossed circles indicate the measurement and the initial instant of the integration time for each of them. Top and bottom show TPI in the case of $630 \pm 30$\;MHz and $354 \pm 17$\;MHz detuning, respectively. \textbf{c)}  corresponds to $g_{HOM}^{(2)}(\tau)$ for couple \#1 when integrated over 3 minutes. d) Excitation spectra as a function of time as in Fig.\ref{fig2}b), obtained in presence of the non-resonant pulsed excitation, for different laser intensities. Statistics on the peak position yields a gaussian-like distribution characterized by $\sigma= 8$;$\,84$;$\,71\,$MHz for illumination power $P=0; 2$ and $5 $ $\mu W$, respectively.}. \label{fig3}
\end{figure}
Finally, HOM interference for photons from remote molecules is obtained with the procedure discussed above. Coincidences are acquired 
in time-tagged mode, allowing for a semi-real time analysis. In panel \ref{fig2}d, an example of the $g^{(2)}_{HOM}(\tau)$ is shown, obtained integrating coincidences within a 30-second time lapse. This interval is found to be a good trade-off, fast enough to minimize the effect of possible frequency shifts due to local charge dynamics, while accumulating sufficient signal with respect to noise. The best fit to the experimental data is obtained with the following function, related to the case of distinct emitters (see black solid line in panel (d)):
\begin{equation}
\begin{split}
g^{(2)}_{HOM}(\tau) = \frac{1}{4} \left ( (g^{(2)}_{HBT,1}(\tau) + g^{(2)}_{HBT,2}(\tau)) + 2\left ( 1-e^{-\frac{T}{T_1}}\right )\sum_{k} e^{-\frac{|\tau + kT|}{T_{1}}} + \right . \\
\left . - 2v|g^{(1)}_{1}(\tau)||g^{(1)}_{2}(\tau)|e^{-2\pi^2(\sigma^2_1 + \sigma^2_2)\tau^2}cos(2\pi\Delta\nu \tau) \right )
\end{split}
\label{HOM}
\end{equation}

where $g^{(1)}_{i}(\tau)=
e^{-\frac{|\tau|}{T_{2,i}}}$, $\sigma^2_i$ is the variance of each molecule central emission frequency over the measurement acquisition time, $\Delta\nu$ is their relative frequency detuning and $v$ is the so called v-factor that is equivalent to a post-selected visibility in zero ($v=1-2g^{(2)}_{HOM}(0)$). 

In this case we found a visibility of about $(40\pm7)\%$ and $v= (96\pm 8)\%$, which is well in line with what found for other solid state quantum emitters under similar pumping conditions \cite{Reindl2017}. Here though, the selected emitters are found and tuned together on the same chip, allowing for a more scalable and integrated configuration. The uncertainty values are estimated by error propagation, based on the integration of poissonian-distributed coincidence counts. 

The attained short-term visibility is well understood on the basis of the previously characterized properties of the individual emitters and on the relative frequency shift. In the reported case for instance, considering $T_1$ and $T_2$ as obtained from the fit and consistent with independent measurements ($T_1=4.25\pm 0.06$ ns \& $FWHM=59\pm4\;$MHz and $T_1=3.88 \pm 0.02$\;ns \&  $FWHM=63\pm2\;$MHz for the two molecules, respectively) and spectral diffusion in the range discussed above, we expect a visibility ranging from $35\%$ to $58\%$, going from a detuning of $50$ to $0$\;MHz. 

Interestingly, thanks to our acquisition mode, we can also appreciate the evolution of the HOM trace in time. The full movie of the HOM trace following Couple \#1 is available as Supplementary Material. Correspondingly, in Fig.\ref{fig3}(a), the visibility is reported as a function of time for $g^{(2)}_{HOM}(\tau)$ integrated over $30$\;s. Each point is shifted by one second and two couples of molecules are analysed in different measurements. Strong variations are observed with hundreds of seconds characteristic time. Negative visibility values are actually all consistent with zero, given the $10\%$ error bar, obtained from poissonian-distributed coincidences.

Real-time HOM data contain many interesting features. In panels \ref{fig3}(b) for instance, two examples are highlighted of measurements whose time stamp corresponds to the different circles in panel (a). In particular, panel \ref{fig3}(b) shows clear interference fringes, whose visibility is accounted for by a v-factor of $0.52 \pm 0.08$ and $0.99 \pm 0.11$ in the top and bottom plot, respectively. In the two cases, the best fit with the function in Eq.\ref{HOM} yields an instantaneous detuning of about $630 \pm 30$ MHz and $354 \pm 17$ MHz. The smallest v factor for the largest detuning is due to the limited temporal resolution of the electronics. Indeed, with a time bin of $500$\;ps, oscillations can be resolved up to a frequency detuning of about $1/(2T_{bin})\simeq 1$\;GHz.

It is worth noticing that, although quantum beats are clearly visible, the data shown in panel (b) correspond to near-zero visibility. The peak suppression is lost already for a detuning $\Delta\nu\simeq 100$\;MHz, when the oscillation period approaches twice the excited state lifetime, i.e. the maximum coherence relaxation time. A smooth drift of the molecule ZPL frequency is hence observed, characterized by typical time scales of the order of hundreds of seconds. HOM measurements allow to detect single-photon frequency detuning smaller than 1 GHz, with 30-seconds integration time and sensitivity close to $20$\;MHz. Detuning values smaller than about $100$\;MHz cannot be clearly determined, as the peak at zero time delay is suppressed and visibility correspondingly increases.

Let us now discuss the impact of a change in the ZPLs' frequency difference, assuming values well beyond $1/T_{bin}$ (temporal resolution) during the acquisition. This is actually relevant when integrating over larger time windows, when a drift of the relative ZPLs can be expected. Fig.\ref{fig3}(c) for instance, represents the HOM signal for Couple \#1 (the same as in Fig.\ref{fig2}(d)), when integrated over 3 minutes. Here indeed, any interference pattern given by the $cos(2\pi\Delta\nu \tau)$ is smeared out, adding up in an increased value at $\tau=0$, hence yielding a reduced v-factor. This phenomenon is clearly visible in the inset to panel \ref{fig3}(c), showing a v-factor of $0.8\pm 0.04$, which is significative smaller than the typical value obtained for single-molecule HOM interference ($v>90\%$). The suppression of the peak at zero time delay is obviously less pronounced than for the single molecule case but also with respect to the short-time HOM trace. It results in a visibility of $0.18\pm 0.04$. 

In order to better understand decoherence mechanisms affecting the long-term HOM signal, excitation-spectra traces have been recorded in presence of the non-resonant excitation laser, for different values of the laser power. Typical results are plotted in Fig.\ref{fig3}(d) for a total interrogation time of about $5\,$min. The presence of the non-resonant excitation laser increases spectral wandering even at low power and determines frequency distributions with standard deviations $\sigma= 8$;$\,84$;$\,71\,$MHz for illumination power values of $P=0; 2$ and $5 $ $\mu W$, respectively (typical values for the HOM measurements). We hence conclude that spectral diffusion is higher in presence of the non-resonant pump beam and affects HOM measurements by a factor that depends on the required integration time.

\section{Conclusion}
In conclusion, in this work we have measured Hong-ou-Mandel interference using distinct optically-tuned molecules on the same chip as single photon sources. The signal to noise was high enough to allow analysing correlations on short time scales and extract information about relative frequency fluctuations of the two molecules' zero phonon lines. We found raw visibilities of the order of $40\%$ and post-selected values up to $97\%$. The tuning method is compatible with pre-structured samples, containing waveguides or cavities and was also demonstrated for multiple emitters simultaneously \cite{Colautti2020b}. The addition of electrodes on a similar geometry could help stabilizing the charge-noise environment, while maintaining the local tuning by optical fields, hence allowing for a higher visibility.
The discussed experiment yields important information in the perspective of scaling up -on integrated platforms- the number of quantum emitters able to generate mutually indistinguishable photons. Overall, our results promote the interest of molecules for the development of photonic quantum simulators and processors. It might be beneficial also for other solid-state systems, suggesting a peculiar optical scheme and data analysis routine for two-photon interference experiments.  

\begin{backmatter}

\bmsection{Acknowledgments}
This project has received funding from the EraNET Cofund Initiatives QuantERA within the European Union's Horizon 2020 research and innovation program grant agreement No. 731473 (project ORQUID), from the FET-OPEN-RIA grant STORMYTUNE (Grant Agree- ment No. 899587) of the European Commission, and from the EMPIR programme (project 17FUN06, SIQUST and 20FUN05 SEQUME), co-financed by the Participating States and from the European Union’s Horizon 2020 research and innovation programme. C.T. would like to thank M. Barbieri for useful discussions.\newline

P.L. and C.T. conceived the research and designed the experiments. All authors performed the experiments; R.D. developed the theoretical model and performed data analysis under the supervision of C.T., who wrote the manuscript with critical feedback from all authors. Figures were prepared by M.C.
\\ The authors declare no conflicts of interest. 

\bmsection{Disclosures}
The authors declare no conflicts of interest.

\bmsection{Data availability} Data underlying the results presented in this paper are not publicly available at this time but may be obtained from the authors upon reasonable request.

\bmsection{Supplemental document}
See Supplement 1 for supporting content. 

\end{backmatter}


\bibliography{publicationsJabRef2}

\end{document}